\documentclass[twocolumn,preprintnumbers,amsmath,amssymb,floatfix,superscriptaddress,nofootinbib]{revtex4}

\usepackage{graphicx}
\usepackage{dcolumn}
\usepackage{bm}
\usepackage{verbatim}
\usepackage{tabularx}
\usepackage{slashed}
\usepackage{cancel}

\def\be{\begin{equation}}
\def\ee{\end{equation}}
\def\bea{\begin{eqnarray}}
\def\eea{\end{eqnarray}}

\newcommand{\vev}[1]{ \left\langle {#1} \right\rangle }

\newcommand{\mr}[1]{\mathrm{#1}}

\begin{document}

\preprint{SLAC-PUB-14834}
\preprint{QMUL-PH-11-18}

\title{Comments on Gaugino Screening}

\author{Timothy Cohen}
\affiliation{Theory Group, SLAC National Accelerator Laboratory,\\ Menlo Park, CA 94025 USA\\\vspace{.15cm}}
\author{Anson Hook}
\affiliation{Theory Group, SLAC National Accelerator Laboratory,\\ Menlo Park, CA 94025 USA\\\vspace{.15cm}}
\author{Brian Wecht}
\affiliation{Center for the Fundamental Laws of Nature, Harvard University,\\ Cambridge, MA 02138 USA\vspace{.15cm}}
\affiliation{Centre for Research in String Theory, Department of Physics,
Queen Mary University of London,
Mile End Road, London E1 4NS, UK}


\begin{abstract}
Gauge mediated models of supersymmetry breaking often exhibit ``gaugino screening," where to leading order in $F$, gaugino masses are unaffected by higher dimensional K\"ahler potential interactions between the supersymmetry breaking spurion and the messengers. We provide a derivation of this phenomenon which utilizes the gaugino counterterm originally proposed in the context of anomaly mediation by Dine and Seiberg.  We argue that this counterterm is present when there are non-zero messenger $F$-terms, and can cancel the leading order Feynman diagram contribution to the gaugino mass.  We provide a nontrivial check of the regulator independence of our results by performing the computation using both dimensional reduction and Pauli-Villars.  This analysis reconciles an apparent contradiction between diagrammatics and analytic continuation into superspace.
\end{abstract}

\maketitle

\section{Introduction}\label{sec:Intro} 

The mediation of supersymmetry breaking (\leavevmode\cancel{SUSY}) to the Standard Model via gauge interactions (for a review see \cite{Giudice:1998bp}) still remains a viable possibility in the early LHC era \cite{Kats:2011qh}, and has many compelling features. First, it provides a calculable framework for coupling (dynamical) \leavevmode\cancel{SUSY} to the Standard Model.  Additionally, the resulting masses are flavor blind, thereby alleviating tension with experimental constraints.  Furthermore, for a large class of models, the sfermion and gaugino masses are of the same parametric size.  This allows one to minimize the fine-tuning required to reproduce the $Z$-boson mass in light of the LEP bound on the Higgs mass.

Unfortunately, there are still several problems with finding realistic gauge mediated models. For example, there are a variety of cases where gauginos can be parametrically lighter than sfermions.  This hierarchy can occur in theories with dynamical \leavevmode\cancel{SUSY} because of the presence of an $R$-symmetry, which exists in generic models where all allowed couplings are non-vanishing \cite{Nelson:1993nf}.  The gauginos $\lambda$ have $R(\lambda)=1$, and hence they cannot obtain a Majorana mass if the hidden sector does not break the $R$-symmetry.

There is another class of models which realize an ``accidental" suppression of gaugino masses, in the sense that the suppression is not the result of a symmetry.  Specifically, assume the mediation is due to the presence of messenger states with mass $M$ charged under the Standard Model.  If the \leavevmode\cancel{SUSY} scale is parametrized by $\sqrt{F}$, one naively expects both gaugino and scalar masses of $\mathcal{O}(F/M)$.  However, when this accidental cancellation occurs, the leading contribution to the gaugino mass is either at three or higher loop order, or at $\mathcal{O}(F^3/M^5)$ \cite{Dumitrescu:2010ha}.

One general set of models with such a suppression was identified by Komargodski and Shih \cite{Komargodski:2009jf}.  They performed a general analysis for theories (with canonical K\"ahler potentials) which could be cast as generalized O'Raifeartaigh models.  In these cases, there is always at least one tree-level flat direction parametrized by a pseudo-modulus.  They demonstrated that in order for the gaugino mass to be non-vanishing at leading order in $F$, there must exist a tachyonic field at some point along this flat direction.  This analysis applies, for example, to models which attempt to break the $R$-symmetry inherent in the original metastable \leavevmode\cancel{SUSY} model of \cite{Intriligator:2006dd} at tree level.

If no tree-level couplings exist between the messengers and the \leavevmode\cancel{SUSY} spurion, it is a logical possibility that such couplings could be generated by loop interactions.  This is the case where the ``gaugino screening" theorem applies \cite{ArkaniHamed:1998kj}. Specifically, this theorem states that higher dimension K\"ahler potential corrections to the messenger sector do not contribute to the gaugino mass at leading order in $F$.  This theorem applies to many models, \emph{e.g.} \cite{Randall:1996zi, Csaki:1997if, Ibe:2007wp, Seiberg:2008qj, Elvang:2009gk, Buican:2008ws, Bazzocchi:2011df}.

The gaugino screening theorem was first proven using analytic continuation into superspace \cite{ArkaniHamed:1998kj}, which derives leading-order effects in $F$ by promoting couplings to superfields with non-zero $F$-term vevs. We review this derivation in Sec.~\ref{sec:OriginalDerivation}.  This theorem was also demonstrated using the language of general gauge mediation in \cite{Dumitrescu:2010ha}.  One feature of both of these proofs is that they rely on arguments which avoid using Feynman diagrams. It is then perhaps surprising that a naive diagrammatic calculation along the lines of \cite{Poppitz:1996xw} completely misses the gaugino screening effect.  For example, it can give non-zero leading order gaugino masses when the other techniques say that they should not exist! Thus, we are left with a natural question of how to demonstrate gaugino screening purely from the diagrammatic point of view. In other words, when can we trust that a diagrammatic calculation gives the correct leading-order gaugino masses?

In this work, we provide the answer to this question. In particular, we show that to give the correct answer, one must include the gaugino counterterm first derived by Dine and Seiberg in \cite{Dine:2007me}. In that context, this term was encountered as an alternative derivation of anomaly mediation; for a similar analysis in AdS space see \cite{Gripaios:2008rg}. We demonstrate that this term is exactly what is needed to cancel the naive leading-order gaugino mass, and is present when there are non-zero messenger $F$-terms. We perform the computation using both dimensional reduction and Pauli-Villars, thus providing a non-trivial check of regulator independence.  Although many of our results may be known in various guises to experts \cite{Dumitrescu:2010ha}, we felt that a detailed exploration of these ideas would make a useful addition to the literature.

The rest of this note is organized as follows.  In the next section, we argue for a simple criterion for vanishing gaugino masses.  In Sec.~\ref{sec:TheGauginoCT} we review the derivation of the gaugino counterterm, how it is sourced by messenger $F$-terms, and the connection with anomaly mediation.  In Sec.~\ref{sec:AToyExample}, we show how the cancellation proceeds in a simple toy model, followed by Sec.~\ref{sec:TheGeneralArgument} where we give our general proof.  For completeness we review the connection with analytic continuation into superspace in Sec.~\ref{sec:OriginalDerivation}.  Finally in Sec.~\ref{sec:CommentsAndConclusions}, we give some general comments and our conclusions.
 
\section{A Simple Criterion}\label{sec:ASimpleCriterion}
\begin{figure}[h!]
\centering
\includegraphics[width=0.35\textwidth]{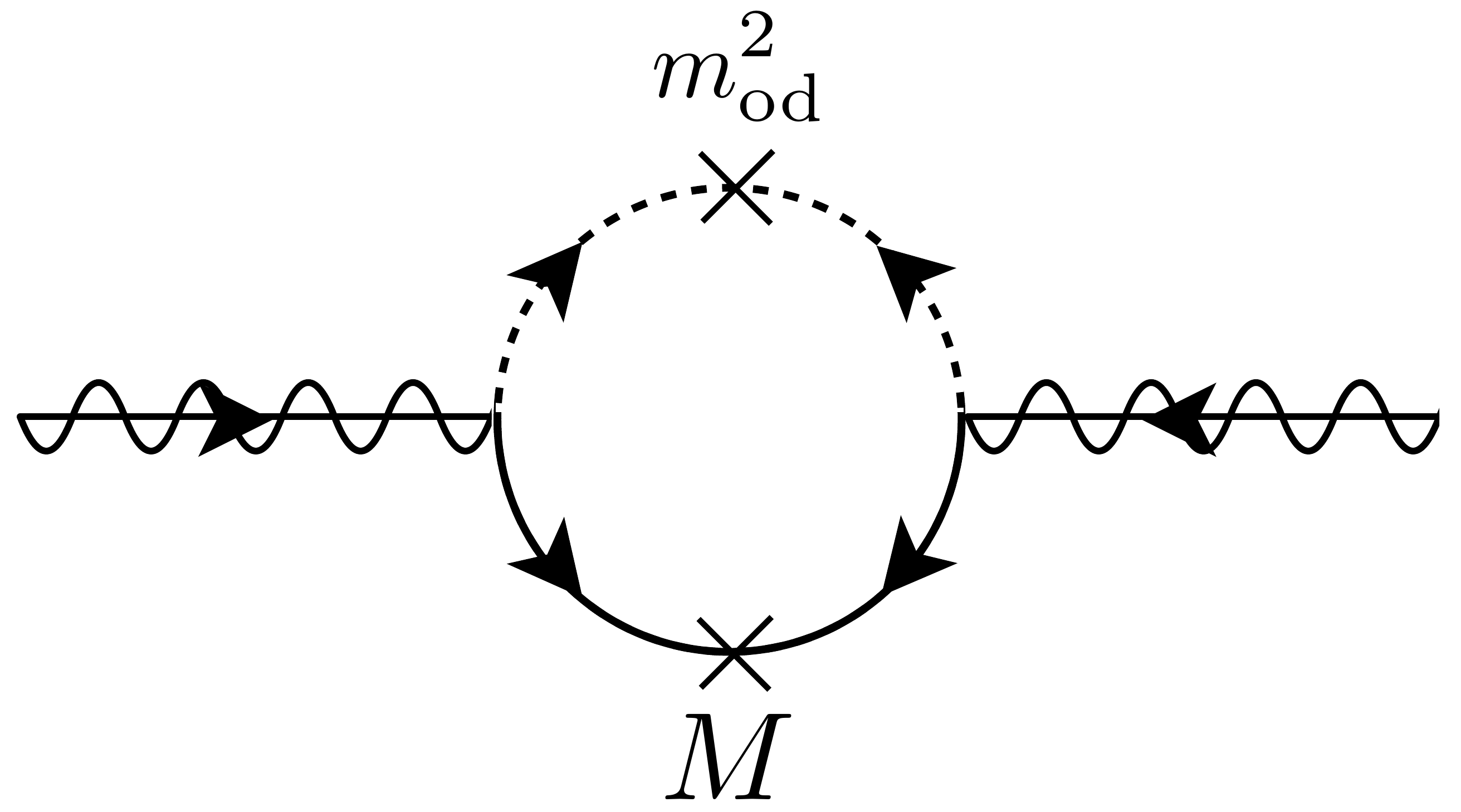}
\caption{The 1-loop contribution to the gaugino mass for models with massive messengers and $m^2_\mr{od}\neq 0$.}
\label{fig:1LoopGauginoMass}
\end{figure}

We begin with a simple criterion for vanishing gaugino masses.  For the model with a single set of messengers, the leading order gaugino mass is given by the 1-loop Feynman diagram of Fig.~\ref{fig:1LoopGauginoMass}, and is
\be\label{logm}
m_{\lambda}^\mr{loop} = \frac{g^2}{16\,\pi^2}\frac{m^2_\mr{od}}M,
\ee
 where $m^2_\mr{od}$ is the \leavevmode\cancel{SUSY} off-diagonal entry in the messenger scalar mass-squared matrix \cite{Poppitz:1996xw}. 

Let $\overline{\Phi}$ and $\Phi$ be vector-like messengers with a supersymmetric mass term $M$.  For simplicity, take the gauge group to be $U(1)$.  We assume that there is no $D$-term breaking, and the scalar messengers have a mass matrix of the form
\be\label{eq:GenericMessengerMatrix}
\mathcal{M}^2 = \left( \begin{array}{cc}
M^2 & M \kappa  \\
M \kappa & M^2 \end{array} \right),
\ee
where $\kappa$ is a mass-dimension one soft supersymmetry breaking parameter whose leading order dependence on \leavevmode\cancel{SUSY} goes as $F/\Lambda$. As will be crucial for our argument, $\Lambda$ is necessarily a scale different from $M$.  A concrete model of such a mass matrix will be provided in Sec.~\ref{sec:AToyExample} below. Regulating this model using Pauli-Villars involves adding additional fields $\Phi'$ and $\overline{\Phi}'$ with mass $\Lambda$ and a mass matrix identical to Eq.~(\ref{eq:GenericMessengerMatrix}), but with $M\rightarrow \Lambda$.

 For the model of Eq.~(\ref{eq:GenericMessengerMatrix}), one obtains a contribution to the gaugino mass from the messengers, given by $m_{\lambda}^{\Phi} = g^2 \kappa /(16\pi^2)$. Additionally, there is a contribution from the Pauli-Villars fields which is identical but with the opposite sign; we will argue that this naive analysis holds in Sec.~\ref{sec:Non-CanonicalBasis:PV}.  This demonstrates that if $m^2_\mr{od} \sim M \kappa$, leading order gaugino masses vanish. 

\section{The Gaugino Counterterm}\label{sec:TheGauginoCT}
Here we will review the derivation of the gaugino counterterm first proposed by Dine and Seiberg in \cite{Dine:2007me}.  Consider the model with massless fields $\Phi$ and $\overline{\Phi}$ which are vector-like under a gauged $U(1)$. These fields will eventually be identified as messengers. On the Higgs branch,  where $\vev{\Phi} = \vev{\overline{\Phi}} = v$, there is a massless composite state which can be described by the gauge invariant combination $\Phi\,\overline{\Phi}$. The tree-level interaction between $\Phi\,\overline{\Phi}$ and the (massive) gauge superfield $V$ goes as $(\Phi\,\overline{\Phi})^2\,V^2$, so the lowest order non-local contribution to the 1PI effective action is at order $V^4$.  Therefore, any contribution to the gaugino mass must be due to local terms.  

In order to derive this contribution explicitly, note the existence of a rescaling chiral anomaly under which $\Phi \rightarrow e^{i \alpha} \Phi$ and $\bar{\Phi} \rightarrow e^{i \alpha} \bar{\Phi}$ while the holomorphic gauge coupling shifts by $\tau \rightarrow \tau + \frac{\alpha}{\pi}$.  Thus the 1PI Lagrangian includes a term
\be\label{eq:PhiPhiBarWWEffAction}
\mathcal{L} \supset \frac{1}{16\, \pi\, i} \int \mr{d}^2 \theta \left(\tau + \frac{i}{2\,\pi} \log\overline{\Phi} \Phi \right ) W^\alpha W_\alpha.
\ee 
This is the effective interaction between the gauge superfield and the massless state.

If one constructs a model where $F_{\Phi}\neq 0$ on the Higgs branch, Eq.~(\ref{eq:PhiPhiBarWWEffAction}) yields the counterterm contribution to the gaugino mass:
\bea
\label{eq:GauginoCT}
m_{\lambda}^\mr{CT} = -\frac{g^2}{16 \,\pi^2} \frac{F_\Phi}{\vev{\Phi}}.
\eea
As we show in Sec.~\ref{sec:TheGeneralArgument}, the proof of the gaugino screening theorem follows from the presence of this counterterm in the effective action. 

We now revisit the simple criterion of the previous section. The mass matrix of Eq.~(\ref{eq:GenericMessengerMatrix}) can be derived from the Lagrangian
\be\label{eq:IntegrateInF}
\mathcal{L} \supset F_{\Phi}\,F^\dag_{\Phi} + M\, \overline{\Phi}\,F_{\Phi}+\kappa\, F_{\Phi}^\dag \Phi + \mbox{h.c.},
\ee
which comes from
\bea
K &=& \Phi^\dag \Phi (1 + \theta^2 \kappa) + \overline{\Phi}^\dag\,\overline{\Phi},\\
W &=& M\,\Phi\,\overline{\Phi}.
\eea
Recall that the gaugino mass vanishes to leading order.

Now consider moving out on the Higgs branch. To do this, we take $M = 0$, which opens the $D$-flat direction $\vev{\Phi} = \vev{\overline{\Phi}} = v$.  From Eq.~(\ref{eq:IntegrateInF}), $F_{\Phi} = - \kappa\,v \neq 0$. Thus $F_\Phi/\langle \Phi \rangle$ is nonzero, and the counterterm does not vanish. As we will see explicitly below, this counterterm will be crucial for reconciling the dimensional reduction (Sec.~\ref{sec:AToyExampleDR}) and Pauli-Villars (Sec.~\ref{sec:Non-CanonicalBasis:PV}) computations, both of which result in vanishing gaugino masses.

There is a confusion now present for models which utilize the gaugino counterterm of Eq.~(\ref{eq:PhiPhiBarWWEffAction}). We have demonstrated the existence of a gaugino counterterm along the Higgs branch. However, if $M \neq 0$, we are no longer allowed to move away from $\langle \Phi \rangle = \langle \overline \Phi \rangle = 0$. So, does this counterterm exist at the origin of moduli space? Note that for all relevant cases, the gaugino mass derived from the counterterm is independent of $v$. However, $\log\Phi \overline{\Phi}$ is singular for $\vev{\Phi} = \vev{\overline{\Phi}} = 0$.  Therefore, one could worry that the extrapolation to the origin is non-trivial.  

One way to establish that the counterterm exists at the origin relies on the regulator fields.  As a result of Eq.~(\ref{eq:PhiPhiBarWWEffAction}), which is independent of regularization scheme, there is a gaugino mass on the Higgs branch.  If regulated with Pauli-Villars, the regulator fields give an additional leading order gauge mediated contribution that does not decouple as their mass is taken to infinity.  Consistency with the 1PI action requires that the Pauli-Villars fields have a counterterm to cancel this 1-loop contribution.  Since the Pauli-Villars fields are pinned to the origin due to their non-zero mass, the counterterm must exist at the origin.

Finally, we discuss why we do not need to include scalar counterterms.  This differs from the supergravity case where, from the point of view of \cite{Dine:2007me}, the anomaly mediated scalar mass counterterms were the result of RG flow.  In models with rigid supersymmetry, RG flow leads to gauge mediated soft masses --- this is the physics of analytic continuation into superspace.  However, the gaugino counterterm is the result of an anomaly, not RG flow.   Therefore it is not seen by gauge mediation alone and must be added by hand.

\subsection{Relation to Anomaly Mediation}
In supergravity, the effective gauge coupling constant $\tau_\mr{eff} = \tau +i \log\Phi \overline{\Phi}/(2\pi)$ can give rise to a gaugino mass.  The Lagrangian contains \cite{Dine:2007me}
\bea
\mathcal{L}  &\supset& \frac{1}{8\,\pi} \, g^{i \bar{i}}\, \frac{\partial \tau_\mr{eff}}{\partial \Phi_i} (D_i W)^* \,\lambda \lambda \nonumber \\
  &=& \frac{\vev{W^*}}{16\, \pi^2\, M_\mr{Pl}^2}  \lambda \lambda+ \cdots,\label{eq:AnomalyMediatedGauginoMass}
\eea
where 
\be
D_i W = \frac{\partial W}{\partial \Phi^i}+\frac{1}{M_\mr{Pl}^2}\frac{\partial K}{\partial \Phi^i} W,
\ee
and we have taken a canonical K\"ahler potential, $g^{i \bar{i}} = \delta^{i \bar{i}}$.  Tuning the cosmological constant to zero enforces that $\vev{W} = m_{3/2}\,M_\mr{Pl}^2$, where $m_{3/2}$ is the gravitino mass.  Plugging this into Eq.~(\ref{eq:AnomalyMediatedGauginoMass}) gives the anomaly mediated contribution to the gaugino mass \cite{Dine:2007me}.

\section{A Toy Example}\label{sec:AToyExample}
In this section we will show how the gaugino counterterm of Eq.~(\ref{eq:GauginoCT}) implies the vanishing of leading order gaugino masses in the ``half-chiral model" of \cite{Dumitrescu:2010ha}:
\bea
K &=& \Phi^\dagger  \,\Phi\left(1 + \frac{X+X^\dagger}{\Lambda}\right) + \overline{\Phi}^\dagger  \,\overline{\Phi}\label{eq:KInNonCanocalBasis} \\
W &=& M\,\overline{\Phi}\,\Phi \label{eq:WInNonCanocalBasis}
\eea
where $\Lambda$ is the cut-off for the model and $X$ is a \leavevmode\cancel{SUSY} spurion, $\vev{X}=\theta^2\,F_X$.  First we will review the argument that the gaugino mass vanishes in the canonically normalized basis \cite{Dumitrescu:2010ha}.  Performing a chiral rotation on $\Phi$ at the superfield level
\be
\Phi \rightarrow \left(1-\frac{X}{\Lambda}\right)\Phi
\ee
gives, to leading order in $1/\Lambda$,
\bea
K &=&  \Phi^\dagger  \,\Phi+ \overline{\Phi}^\dagger  \,\overline{\Phi} \\
W &=& M\,  \overline{\Phi}\,\Phi - \left(\frac{M\,X}{\Lambda}\right) \overline{\Phi}\,\Phi\\
\mathcal{L}&\supset& \int \frac{\mr{d}^2 \theta}{16 \pi i}\left(\tau +\frac{i}{2\pi} \log\left(1-\frac{X}{\Lambda} \right)  \right )  W^\alpha W_\alpha,
\eea
where the $\log (1-X/\Lambda)$ term results from the anomalous chiral rotation of $\Phi$.  The messenger mass matrix is given by
\be\label{eq:HalfChiralMessengerMatrix}
\mathcal{M}^2_\mr{half-chiral} = \left( \begin{array}{cc}
M^2 & \frac{M\,F_X}{\Lambda}  \\
 \frac{M\,F_X}{\Lambda} & M^2 \end{array} \right),
\ee
In this basis, the off-diagonal mass appears as an independent parameter.  Therefore, even though $M$ appears in $m_\mr{od}^2$, there is {\it a priori} no reason for the supersymmetric messenger mass and this parameter to be related.  This is relevant if one wishes to regulate the theory using Pauli-Villars.  The diagonal element of the Pauli-Villars mass matrix is $M^2_\mr{PV}\rightarrow \infty$ while $m^2_\mr{od} = M\,F_X/\Lambda$.  This implies that the Pauli-Villars fields will \emph{not} make a contribution to the gaugino mass in this basis.

Evaluating the diagram in Fig.~\ref{fig:1LoopGauginoMass} using either dimensional reduction or Pauli-Villars gives a non-zero gaugino mass contribution $g^2 F_X /(16\pi^2 \Lambda)$.  Note that since $K$ is canonical, $F_{\Phi} = 0$ and Eq.~(\ref{eq:GauginoCT}) does not contribute to the gaugino mass.  However, there is a contribution to $m_{\lambda}$ from the $\log(1-X/\Lambda)$ term: $-g^2 F_X /(16\pi^2 \Lambda)$.  Clearly this cancels against the 1-loop diagram.  The gaugino mass vanishes to leading order in $F_X$.

As we now discuss, the computation is more subtle when we do not canonically normalize $\Phi$.  As opposed to the case just discussed, the precise way to see that the gaugino mass vanishes at $\mathcal{O}(F)$ is different for dimensional reduction and Pauli-Villars.

\subsection{Non-Canonical Basis: Dimensional Reduction}\label{sec:AToyExampleDR}
In the non-canonical basis of Eq.~(\ref{eq:KInNonCanocalBasis}), the messenger scalar mass-squared matrix is also given by Eq.~(\ref{eq:HalfChiralMessengerMatrix}).  Then the 1-loop contribution, regulated using dimensional reduction, to the gaugino masses is given by $-g^2 F_X /(16 \pi^2 \Lambda)$.  In this basis, $F_{\Phi} \neq 0$ on the Higgs branch.  This gives a counterterm contribution of $g^2 F_X /(16 \pi^2 \Lambda)$, which precisely cancels the 1-loop diagram.  We see that leading order gaugino masses vanish in this model. 

\subsection{Non-Canonical Basis: Pauli-Villars}\label{sec:Non-CanonicalBasis:PV}
In order to regularize the theory we add the Pauli-Villars fields $\Phi'$ and $\overline{\Phi}'$.  The K\"ahler interactions for these fields are then identical to those for $\Phi$ and $\overline{\Phi}$.  Their mass is given by $M_\mr{PV}$, which will be sent to infinity at the end of the computation.     

There are counterterms for both the $(\Phi,\,\overline{\Phi})$ and $(\Phi',\,\overline{\Phi}')$ fields.  The Pauli-Villars fields have opposite sign $F$-terms, so these counterterms cancel.\footnote{We thank D. Shih for pointing this out to us.}  This shows that when using the Pauli-Villars regularization scheme, the counterterm has no effect.  Hence, the naive argument presented in Sec.~\ref{sec:ASimpleCriterion} follows, and the leading order gaugino masses vanish.

\section{The General Argument}\label{sec:TheGeneralArgument}
So far, we have seen how the counterterm implies vanishing gaugino masses for a specific case.  Here we will provide an analysis relevant for a wide class of models.  Again, let $\Phi$ and $\overline{\Phi}$ be messengers, and go to a basis where all of the \leavevmode\cancel{SUSY} is parametrized by $\vev{X}=F_X\, \theta^2$.  Then take the K\"ahler potential and superpotential to be
\bea
K &=& \Phi^{\dag}\Phi\left(1+f_K(X,X^{\dag}, \dots)\right)\nonumber\\
&+&\overline{\Phi}^{\dag}\overline{\Phi}\left(1+\overline{f}_K(X, X^{\dag}, \dots)\right)+\dots, \label{eq:GenericK}\\
W &=& (M+f_W(X,\dots))\, \Phi\,\overline{\Phi}+\dots,\label{eq:GenericW}
\eea
where $f_K$ and $\overline{f}_K$ are generic real dimensionless superfield functions and $f_W$ is a generic holomorphic superfield function with mass dimension one. For models with multiple messengers, we take the K\"ahler potential to be diagonal.\footnote{If the model is written in a basis with a non-diagonal K\"ahler potential, one can always go to a basis where $K$ is diagonal by acting with the appropriate unitary rotation matrix $U$, $\overline{U}$ on the messenger fields $\Phi$, $\overline{\Phi}$.  This in turn rotates the superpotential couplings.  Then the gaugino mass derived from the rotated superpotential is proportional to $\partial_X \log\det \overline{U} (M+f_W)\, U = \partial_X \log\det (M+f_W)$  \cite{Giudice:1997ni}.  Hence, the gaugino masses are unaffected by non-diagonal K\"ahler potential interactions and our general argument holds.}  We only need to consider terms that are quadratic in $\Phi$ and $\overline{\Phi}$ to determine the leading order behavior of the gaugino masses. We neglect K\"ahler potential terms of the form $g(X, X^\dag,\dots) \Phi \overline \Phi$, since we can always absorb the effect of such operators into a superpotential mass term, to leading order in $F$ \cite{Giudice:1988yz}.  In this model, the messenger $F$-terms are given by
\bea
 F_{\Phi} &=& \frac{-1}{1+f_K} \left(\Phi\, F_X \frac{\partial f_K}{\partial X}+\left(M^\dag+f_W^\dag\right)\overline{\Phi}^\dag\right),\label{eq:GeneralFPhi} \\
F_{\overline{\Phi}} &=& \frac{-1}{1+\overline{f}_K} \left(\overline{\Phi}\, F_X \frac{\partial \overline{f}_K}{\partial X}+\left(M^\dag+f_W^\dag\right)\Phi^\dag\right).\label{eq:GeneralFPhiBar}
\eea
The potential is
\bea
V&=&\frac{1}{1+f_K}\left|\Phi\, F_X\, \frac{\partial f_K}{\partial X}+\left(M^\dag+f_W^\dag\right)\,\overline{\Phi}^\dag\right|^2 \nonumber \\
&+&\frac{1}{1+\overline{f}_K}\left|\overline{\Phi}\, F_X\, \frac{\partial \overline{f}_K}{\partial X}+\left(M^\dag+f_W^\dag\right)\,\Phi^\dag\right|^2 \nonumber \\
&-& \left(\frac{\partial f_W}{\partial X}F_X\, \Phi\,\overline{\Phi} +\mr{h.c.}\right).
\eea
which, after canonically normalizing the component fields, gives
\be
m_{\lambda}^\mr{loop} = -\frac{g^2 F_X }{16 \pi^2} \left(\frac{1}{1+f_K}\frac{\partial f_K}{\partial X}+\frac{1}{1+\overline{f}_K}\frac{\partial \overline{f}_K}{\partial X} 
- \frac{\partial f_W}{\partial X} \right).
\ee
There is also the gaugino counterterm contribution which is present due to the non-zero messenger $F$-terms given in Eqs.~(\ref{eq:GeneralFPhi}) and (\ref{eq:GeneralFPhiBar}):
\be\label{eq:GauginoMassCTGeneral}
m_{\lambda}^\mr{CT} = \frac{g^2 F_X}{16 \pi^2} \left(\frac{1}{1+f_K} \frac{\partial f_K}{\partial X}+\frac{1}{1+\overline{f}_K} \frac{\partial \overline{f}_K}{\partial X}\right).
\ee
We see that the leading order contribution from the K\"ahler potential cancels regardless of the form of $f_K$ and $\overline{f}_K$ while the contribution from the superpotential survives.

This proves the gaugino screening theorem in generality.  If \leavevmode\cancel{SUSY} does not communicate with the messengers at tree level (\emph{e.g.}~via a superpotential interaction $X\, \Phi\, \overline{\Phi}$), then the coupling between the messengers and the spurion can be put in the form of Eqs.~(\ref{eq:GenericK}) and (\ref{eq:GenericW}).  Therefore, K\"ahler potential corrections involving the messengers and the hidden section do not induce leading order gaugino masses.  

It is straightforward to see why the cancellation does not generically hold to all orders in $F_X$.   The 1-loop diagram generates contributions which are higher order in $F_X/M^2$ while the counterterm contribution is given exactly by Eq.~(\ref{eq:GauginoMassCTGeneral}).

\section{Derivation by Analytic Continuation into Superspace}\label{sec:OriginalDerivation}
Another way of computing the soft spectrum in gauge mediated models utilizes an analytic continuation into superspace of the running gauge coupling into superspace.  The original work \cite{Giudice:1997ni} provided a systemic procedure for extracting the leading order contribution to the soft-masses.  In \cite{ArkaniHamed:1998kj} this was extended to include higher loop effects.  In particular, by including a threshold correction from the wavefunction renormalization of the messengers, they were able to prove the gaugino screening theorem.

For completeness, we will summarize the proof of \cite{ArkaniHamed:1998kj} here.  The gaugino mass is determined by the $\theta^2$ component of a real superfield,
\bea
R(\mu) &=& S(\mu) + S^\dagger(\mu) + \frac{T_G}{8 \pi^2} \log\left( S(\mu) + S^\dagger(\mu) \right)\nonumber \\ 
&-& \sum_r \frac{T_r}{8 \pi^2} \log (Z_r(\mu)) + \mbox{2-loop}.\label{eq:R}
\eea
$Z_r$ is the wavefunction renormalization of all fields charged under the gauge group.  $S(\mu)$ is the holomorphic gauge coupling and is determined exactly by 
\bea
S(\mu) = S(\Lambda) + \frac{b_0}{16 \pi^2} \log\left(\frac{\mu}{\Lambda}\right)- \frac{T_\Phi}{16 \pi^2} \log\left(\frac{M_\Phi}{\Lambda}\right),
\eea
where $b_0$ is the one-loop beta function without the messenger fields $\Phi$, $M_\Phi$ is the physical mass of the messengers, and we evaluated this coupling at a scale $\mu < M_\Phi < \Lambda$. To leading order in $g^2$, the $X$-dependent part of the real gauge coupling constant is 
\bea
R(\mu) = -\frac{T_\Phi}{8 \pi^2} \log\left(\frac{M_\Phi Z_\Phi(M_\Phi)}{\Lambda}\right) + \dots
\eea
Given the superpotential messenger mass $M$, the physical mass is given by
\be
M_\Phi = \frac{M}{Z_\Phi(M_\Phi)}.
\ee

This relationship, along with Eq.~(\ref{eq:R}), implies that the dependence of the gaugino mass on the wave function of the messengers cancels at leading order in $F$ --- any non-vanishing leading order gaugino mass is completely determined by superpotential couplings.  This is how the gaugino screening theorem manifests itself using analytic continuation methods \cite{ArkaniHamed:1998kj}.  We note that this analysis avoids any reference to Feynman diagrams or counterterms.

\section{Comments and Conclusions}\label{sec:CommentsAndConclusions}
In this note we have given an alternative derivation of the gaugino screening theorem.  As opposed to the original result which relied on analytic continuation into superspace, we have shown how this effect is realized at the level of Feynman diagrams.  In particular, the presence of non-zero messenger $F$-terms necessitates the inclusion of gaugino counterterms.  These finite counterterms are responsible for the gaugino screening effect.  This provides an explicit class of models where one must be careful in performing diagrammatic gauge mediation analyses. 

We conclude with some comments on the UV interpretation of the gaugino counterterm.\footnote{We are indebted to N.~Seiberg for explaining this point to us.}  Intuitively, this non-decoupling effect is special due to its relation with the chiral anomaly.  In \cite{Buican:2008ws}, a generic criterion for UV sensitivity in models of gauge mediation was given.  These authors proposed a simple model which exhibits vanishing leading order gaugino masses (their example is similar to the half-chiral model discussed in Sec.~\ref{sec:AToyExample} above).  In the IR description of their model, there are non-zero messenger $F$-terms and therefore the gaugino counterterm is non-zero and completely cancels the leading order gaugino mass.
From the UV completion, the cancellation for the leading order gaugino mass can be seen explicitly utilizing a Feynman diagram computation.  This provides an explicit case where the counterterm can be understood in terms of contact contributions from a UV threshold.  This makes it clear that one must include the gaugino counterterm at low energies to maintain the consistency of the model.

\begin{acknowledgments}
We would like to thank Michael Dine, Michael Peskin, Aaron Pierce, David Morrissey, Nathan Seiberg, David Shih, Tomer Volansky, and Jay Wacker for discussions.
TC  and AH are supported by the US Department of Energy under contract number DE-AC02-76SF00515.  
BW is supported by the Fundamental Laws Initiative of the Center for
the Fundamental Laws of Nature, Harvard University and the STFC Standard Grant ST/J000469/1 ``String Theory, Gauge Theory and Duality."
\end{acknowledgments}

\newpage

\bibliography{GauginoScreening}

\begin{thebibliography}{19}
\expandafter\ifx\csname natexlab\endcsname\relax\def\natexlab#1{#1}\fi
\expandafter\ifx\csname bibnamefont\endcsname\relax
  \def\bibnamefont#1{#1}\fi
\expandafter\ifx\csname bibfnamefont\endcsname\relax
  \def\bibfnamefont#1{#1}\fi
\expandafter\ifx\csname citenamefont\endcsname\relax
  \def\citenamefont#1{#1}\fi
\expandafter\ifx\csname url\endcsname\relax
  \def\url#1{\texttt{#1}}\fi
\expandafter\ifx\csname urlprefix\endcsname\relax\def\urlprefix{URL }\fi
\providecommand{\bibinfo}[2]{#2}
\providecommand{\eprint}[2][]{\url{#2}}

\bibitem[{\citenamefont{Giudice and Rattazzi}(1999)}]{Giudice:1998bp}
\bibinfo{author}{\bibfnamefont{G.}~\bibnamefont{Giudice}} \bibnamefont{and}
  \bibinfo{author}{\bibfnamefont{R.}~\bibnamefont{Rattazzi}},
  \bibinfo{journal}{Phys.Rept.} \textbf{\bibinfo{volume}{322}},
  \bibinfo{pages}{419} (\bibinfo{year}{1999}), \eprint{hep-ph/9801271}.

\bibitem[{\citenamefont{Kats et~al.}(2011)\citenamefont{Kats, Meade, Reece, and
  Shih}}]{Kats:2011qh}
\bibinfo{author}{\bibfnamefont{Y.}~\bibnamefont{Kats}},
  \bibinfo{author}{\bibfnamefont{P.}~\bibnamefont{Meade}},
  \bibinfo{author}{\bibfnamefont{M.}~\bibnamefont{Reece}}, \bibnamefont{and}
  \bibinfo{author}{\bibfnamefont{D.}~\bibnamefont{Shih}}
  (\bibinfo{year}{2011}), \eprint{1110.6444}.

\bibitem[{\citenamefont{Nelson and Seiberg}(1994)}]{Nelson:1993nf}
\bibinfo{author}{\bibfnamefont{A.~E.} \bibnamefont{Nelson}} \bibnamefont{and}
  \bibinfo{author}{\bibfnamefont{N.}~\bibnamefont{Seiberg}},
  \bibinfo{journal}{Nucl. Phys.} \textbf{\bibinfo{volume}{B416}},
  \bibinfo{pages}{46} (\bibinfo{year}{1994}), \eprint{hep-ph/9309299}.

\bibitem[{\citenamefont{Dumitrescu et~al.}(2010)\citenamefont{Dumitrescu,
  Komargodski, Seiberg, and Shih}}]{Dumitrescu:2010ha}
\bibinfo{author}{\bibfnamefont{T.~T.} \bibnamefont{Dumitrescu}},
  \bibinfo{author}{\bibfnamefont{Z.}~\bibnamefont{Komargodski}},
  \bibinfo{author}{\bibfnamefont{N.}~\bibnamefont{Seiberg}}, \bibnamefont{and}
  \bibinfo{author}{\bibfnamefont{D.}~\bibnamefont{Shih}},
  \bibinfo{journal}{JHEP} \textbf{\bibinfo{volume}{1005}}, \bibinfo{pages}{096}
  (\bibinfo{year}{2010}), \eprint{1003.2661}.

\bibitem[{\citenamefont{Komargodski and Shih}(2009)}]{Komargodski:2009jf}
\bibinfo{author}{\bibfnamefont{Z.}~\bibnamefont{Komargodski}} \bibnamefont{and}
  \bibinfo{author}{\bibfnamefont{D.}~\bibnamefont{Shih}},
  \bibinfo{journal}{JHEP} \textbf{\bibinfo{volume}{0904}}, \bibinfo{pages}{093}
  (\bibinfo{year}{2009}), \eprint{0902.0030}.

\bibitem[{\citenamefont{Intriligator et~al.}(2006)\citenamefont{Intriligator,
  Seiberg, and Shih}}]{Intriligator:2006dd}
\bibinfo{author}{\bibfnamefont{K.~A.} \bibnamefont{Intriligator}},
  \bibinfo{author}{\bibfnamefont{N.}~\bibnamefont{Seiberg}}, \bibnamefont{and}
  \bibinfo{author}{\bibfnamefont{D.}~\bibnamefont{Shih}},
  \bibinfo{journal}{JHEP} \textbf{\bibinfo{volume}{0604}}, \bibinfo{pages}{021}
  (\bibinfo{year}{2006}), \eprint{hep-th/0602239}.

\bibitem[{\citenamefont{Arkani-Hamed et~al.}(1998)\citenamefont{Arkani-Hamed,
  Giudice, Luty, and Rattazzi}}]{ArkaniHamed:1998kj}
\bibinfo{author}{\bibfnamefont{N.}~\bibnamefont{Arkani-Hamed}},
  \bibinfo{author}{\bibfnamefont{G.~F.} \bibnamefont{Giudice}},
  \bibinfo{author}{\bibfnamefont{M.~A.} \bibnamefont{Luty}}, \bibnamefont{and}
  \bibinfo{author}{\bibfnamefont{R.}~\bibnamefont{Rattazzi}},
  \bibinfo{journal}{Phys.Rev.} \textbf{\bibinfo{volume}{D58}},
  \bibinfo{pages}{115005} (\bibinfo{year}{1998}), \eprint{hep-ph/9803290}.

\bibitem[{\citenamefont{Randall}(1997)}]{Randall:1996zi}
\bibinfo{author}{\bibfnamefont{L.}~\bibnamefont{Randall}},
  \bibinfo{journal}{Nucl.Phys.} \textbf{\bibinfo{volume}{B495}},
  \bibinfo{pages}{37} (\bibinfo{year}{1997}), \eprint{hep-ph/9612426}.

\bibitem[{\citenamefont{Csaki et~al.}(1998)\citenamefont{Csaki, Randall, and
  Skiba}}]{Csaki:1997if}
\bibinfo{author}{\bibfnamefont{C.}~\bibnamefont{Csaki}},
  \bibinfo{author}{\bibfnamefont{L.}~\bibnamefont{Randall}}, \bibnamefont{and}
  \bibinfo{author}{\bibfnamefont{W.}~\bibnamefont{Skiba}},
  \bibinfo{journal}{Phys.Rev.} \textbf{\bibinfo{volume}{D57}},
  \bibinfo{pages}{383} (\bibinfo{year}{1998}), \eprint{hep-ph/9707386}.

\bibitem[{\citenamefont{Ibe et~al.}(2007)\citenamefont{Ibe, Nakayama, and
  Yanagida}}]{Ibe:2007wp}
\bibinfo{author}{\bibfnamefont{M.}~\bibnamefont{Ibe}},
  \bibinfo{author}{\bibfnamefont{Y.}~\bibnamefont{Nakayama}}, \bibnamefont{and}
  \bibinfo{author}{\bibfnamefont{T.}~\bibnamefont{Yanagida}},
  \bibinfo{journal}{Phys.Lett.} \textbf{\bibinfo{volume}{B649}},
  \bibinfo{pages}{292} (\bibinfo{year}{2007}), \eprint{hep-ph/0703110}.

\bibitem[{\citenamefont{Seiberg et~al.}(2008)\citenamefont{Seiberg, Volansky,
  and Wecht}}]{Seiberg:2008qj}
\bibinfo{author}{\bibfnamefont{N.}~\bibnamefont{Seiberg}},
  \bibinfo{author}{\bibfnamefont{T.}~\bibnamefont{Volansky}}, \bibnamefont{and}
  \bibinfo{author}{\bibfnamefont{B.}~\bibnamefont{Wecht}},
  \bibinfo{journal}{JHEP} \textbf{\bibinfo{volume}{0811}}, \bibinfo{pages}{004}
  (\bibinfo{year}{2008}), \eprint{0809.4437}.

\bibitem[{\citenamefont{Elvang and Wecht}(2009)}]{Elvang:2009gk}
\bibinfo{author}{\bibfnamefont{H.}~\bibnamefont{Elvang}} \bibnamefont{and}
  \bibinfo{author}{\bibfnamefont{B.}~\bibnamefont{Wecht}},
  \bibinfo{journal}{JHEP} \textbf{\bibinfo{volume}{0906}}, \bibinfo{pages}{026}
  (\bibinfo{year}{2009}), \eprint{0904.4431}.

\bibitem[{\citenamefont{Buican et~al.}(2009)\citenamefont{Buican, Meade,
  Seiberg, and Shih}}]{Buican:2008ws}
\bibinfo{author}{\bibfnamefont{M.}~\bibnamefont{Buican}},
  \bibinfo{author}{\bibfnamefont{P.}~\bibnamefont{Meade}},
  \bibinfo{author}{\bibfnamefont{N.}~\bibnamefont{Seiberg}}, \bibnamefont{and}
  \bibinfo{author}{\bibfnamefont{D.}~\bibnamefont{Shih}},
  \bibinfo{journal}{JHEP} \textbf{\bibinfo{volume}{0903}}, \bibinfo{pages}{016}
  (\bibinfo{year}{2009}), \eprint{0812.3668}.

\bibitem[{\citenamefont{Bazzocchi and Monaco}(2011)}]{Bazzocchi:2011df}
\bibinfo{author}{\bibfnamefont{F.}~\bibnamefont{Bazzocchi}} \bibnamefont{and}
  \bibinfo{author}{\bibfnamefont{M.}~\bibnamefont{Monaco}}
  (\bibinfo{year}{2011}), \eprint{1111.1122}.

\bibitem[{\citenamefont{Poppitz and Trivedi}(1997)}]{Poppitz:1996xw}
\bibinfo{author}{\bibfnamefont{E.}~\bibnamefont{Poppitz}} \bibnamefont{and}
  \bibinfo{author}{\bibfnamefont{S.~P.} \bibnamefont{Trivedi}},
  \bibinfo{journal}{Phys.Lett.} \textbf{\bibinfo{volume}{B401}},
  \bibinfo{pages}{38} (\bibinfo{year}{1997}), \eprint{hep-ph/9703246}.

\bibitem[{\citenamefont{Dine and Seiberg}(2007)}]{Dine:2007me}
\bibinfo{author}{\bibfnamefont{M.}~\bibnamefont{Dine}} \bibnamefont{and}
  \bibinfo{author}{\bibfnamefont{N.}~\bibnamefont{Seiberg}},
  \bibinfo{journal}{JHEP} \textbf{\bibinfo{volume}{0703}}, \bibinfo{pages}{040}
  (\bibinfo{year}{2007}), \eprint{hep-th/0701023}.

\bibitem[{\citenamefont{Gripaios et~al.}(2009)\citenamefont{Gripaios, Kim,
  Rattazzi, Redi, and Scrucca}}]{Gripaios:2008rg}
\bibinfo{author}{\bibfnamefont{B.}~\bibnamefont{Gripaios}},
  \bibinfo{author}{\bibfnamefont{H.~D.} \bibnamefont{Kim}},
  \bibinfo{author}{\bibfnamefont{R.}~\bibnamefont{Rattazzi}},
  \bibinfo{author}{\bibfnamefont{M.}~\bibnamefont{Redi}}, \bibnamefont{and}
  \bibinfo{author}{\bibfnamefont{C.}~\bibnamefont{Scrucca}},
  \bibinfo{journal}{JHEP} \textbf{\bibinfo{volume}{0902}}, \bibinfo{pages}{043}
  (\bibinfo{year}{2009}), \eprint{0811.4504}.

\bibitem[{\citenamefont{Giudice and Rattazzi}(1998)}]{Giudice:1997ni}
\bibinfo{author}{\bibfnamefont{G.}~\bibnamefont{Giudice}} \bibnamefont{and}
  \bibinfo{author}{\bibfnamefont{R.}~\bibnamefont{Rattazzi}},
  \bibinfo{journal}{Nucl.Phys.} \textbf{\bibinfo{volume}{B511}},
  \bibinfo{pages}{25} (\bibinfo{year}{1998}), \eprint{hep-ph/9706540}.

\bibitem[{\citenamefont{Giudice and Masiero}(1988)}]{Giudice:1988yz}
\bibinfo{author}{\bibfnamefont{G.}~\bibnamefont{Giudice}} \bibnamefont{and}
  \bibinfo{author}{\bibfnamefont{A.}~\bibnamefont{Masiero}},
  \bibinfo{journal}{Phys.Lett.} \textbf{\bibinfo{volume}{B206}},
  \bibinfo{pages}{480} (\bibinfo{year}{1988}).

\end{thebibliography}

\end{document}